\documentclass[dvipdfmx]{jpsj3}

\usepackage{txfonts}
\usepackage{bm}

\newcommand{\kB}{k_\mathrm{B}}
\newcommand{\bmxi}{\bm{\xi}}

\title{Dissipative Heat Decomposition in Stochastic Energetics: Implication of the Instantaneous Diffusion Coefficient in Nonequilibrium Steady States}

\author{Soya Shinkai\thanks{soya@hiroshima-u.ac.jp}}

\inst{Research Center for the Mathematics on Chromatin Live Dynamics (RcMcD), Hiroshima University, 739-8530, Japan}

\abst{
We  give a decomposition expression for dissipative heat using the instantaneous diffusion coefficient in a nonequilibrium steady state.
The dissipative heat can be expressed using three diffusion coefficients: instantaneous, equilibrium, and drift.
An experimental application of the decomposition expression permits us to evaluate the heat dissipation rate from single-trajectory data only.
We also numerically demonstrate this method.
}

\begin{document}

\maketitle

Recent developments in microscopy and labeling techniques have enabled us to observe the motion of tiny single particles in living cells with high space and time resolutions.
In analyzing the tracked trajectory data, $\{ x_0, x_1, \cdots, x_n, \cdots, x_{N-1} \}$,
for which the time resolution is $\Delta t$ and $x_n \equiv x(t_n)$ for $t_n \equiv n \, \Delta t$,
as much information as possible is expected to be extracted from the data.
For diffusion in living cells, the most common and conventional analytical method is the mean-squared displacement (MSD) analysis.
Not only the diffusion coefficient but also the drift velocity and a characteristic diffusion length within a limited area can be estimated \cite{Kusumi1993}.
However, as the MSD analysis is based on stationary stochastic processes, we cannot examine its relationship to nonequilibrium biological activity in living cells.

Meanwhile, the development of stochastic energetics has enabled us to energetically describe and estimate phenomena in the fluctuating world \cite{Sekimoto1997, Sekimoto2010}.
Violations of the fluctuation-dissipation relation in nonequilibrium steady states (NESSs) can be expressed in terms of the heat dissipation rate into the thermal environment without a prior knowledge about energy potential \cite{Harada2005}, and this theoretical expression has been applied to an evaluation of the single-molecule energetics of a rotary molecular motor, $\mathrm{F}_1$-ATPase \cite{Toyabe2010}.
However, in general, measurement of the mechanical responses of molecules is a difficult task.
In order to overcome these difficulties, Toyabe et al. developed an analytical method to evaluate the potential profile at each chemical state from a single-molecule trajectory only, without the need for response measurement, and applied it to $\mathrm{F}_1$-ATPase \cite{Toyabe2012}.
In addition, we have also suggested an analytical method for a single-diffusion trajectory based on stochastic energetics \cite{ShinkaiTogashi2014}, which can be used to estimate the enhanced motion due to active diffusion in living cells \cite{Activediffusion}. Specifically, the instantaneous diffusion coefficient (IDC), $D_\mathrm{inst}(t_n) \equiv (\Delta x_n)^2 / (2 \Delta t)$, where $\Delta x_n \equiv x_{n+1} - x_n$, is an energetically well-defined physical quantity for active diffusion trajectories.
However, when the response takes the form of contributions to a potential, the implication of the IDC remains unclear.


Here, we study a system described by an overdamped Langevin equation
\begin{equation}\label{eq:1}
	\gamma \frac{dx(t)}{dt} = - U'(x) + \sqrt{2 \gamma \kB T} \, \hat{\xi}(t),
\end{equation}
where $\gamma$ is the friction coefficient,
$U(x)$ is an arbitrary potential including NESS cases,
$T$ is the environment temperature,
and $\hat{\xi}(t)$ stands for the Gaussian white noise, satisfying $\langle \hat{\xi}(t) \rangle = 0$ and $\langle \hat{\xi}(t) \hat{\xi}(s) \rangle = \delta(t-s)$.
The $\langle \cdot \rangle$ symbol denotes the ensemble average.
Let us consider the integral of Eq.~(\ref{eq:1}) within the time interval, $[t_n, t_{n+1})$, which corresponds to the displacement, $\Delta x_n$, in experiments with time resolution, $\Delta t$.
Since the stochastic differential equation (SDE) corresponding to the Langevin equation (Eq. (\ref{eq:1})) should be interpreted as Stratonovich-type in stochastic energetics \cite{Sekimoto2010},
the integral is expressed as
\begin{equation}\label{eq:2}
	\Delta x_n = - \frac{U'(\tilde{x}_n)}{\gamma} \Delta t
	+ \sqrt{\frac{2 \kB T \Delta t}{\gamma}} \bmxi_n + o(\Delta t),
\end{equation}
where $\tilde{x}_n \equiv (x_n + x_{n+1}) / 2$,
$\{ \bmxi_n \}$ is a sequence of the Gaussian random variables satisfying $\langle \bmxi_n \rangle = 0$ and $\langle \bmxi_n \bmxi_m \rangle = \delta_{nm}$,
and $o(\Delta t)$ denotes terms of order greater than $\Delta t$.

Next, let us consider the dissipative heat into the thermal environment, $\Delta Q_n$, during the time interval, $[t_n, t_{n+1})$.
According to stochastic energetics \cite{Sekimoto2010},
the heat at time $t$ is defined as $d'Q(t) \equiv \left( \gamma \dot{x}(t) - \sqrt{2 \gamma \kB T} \, \hat{\xi}(t) \right) \circ dx(t)$,
where $\circ$ represents Stratonovich multiplication \cite{Gardiner2009}.
Using Eq.~(\ref{eq:1}), this is written as $d'Q(t) = - U'(x) \circ dx(t)$ \cite{Comment1}.
Therefore, for discrete data, $\Delta Q_n$ corresponds to $ - U'(\tilde{x}_n) \Delta x_n$.
Since Eq.~(\ref{eq:2}) can be transformed into the identical relation
\begin{equation}\label{eq:3}
	\left( \Delta x_n + \frac{U'(\tilde{x}_n)}{\gamma} \Delta t \right)^2 \Bigg/ (2 \Delta t)
	= \frac{\kB T}{\gamma} \bmxi_n^2 + o(\Delta t),
\end{equation}
we can obtain the following expression
\begin{equation}\label{eq:4}
	\frac{U'(\tilde{x}_n)}{\gamma} \Delta x_n =
	- \frac{(\Delta x_n)^2 + \left( U'(\tilde{x}_n) \Delta t / \gamma \right)^2}{2 \Delta t}
	+ \frac{\kB T}{\gamma} \bmxi_n^2 + o(\Delta t).
\end{equation}
This equation leads to an expression for the dissipative heat where
\begin{equation}\label{eq:5}
	\Delta Q_n = \gamma \left[
	\frac{(\Delta x_n)^2 + \left( U'(\tilde{x}_n) \Delta t / \gamma \right)^2}{2 \Delta t}
	- \frac{\kB T}{\gamma} \bmxi_n^2 \right] + o(\Delta t).
\end{equation}
Since the average of Eq.~(\ref{eq:2}) at a given initial position $z$ becomes  $\langle \Delta x_n \rangle_z = - U'(z) \Delta t / \gamma + o(\Delta t)$ \cite{Comment2},
we can obtain
\begin{equation}\label{eq:6}
	\Delta Q_n = \gamma \left[
	\frac{(\Delta x_n)^2}{2 \Delta t} + \frac{\langle \Delta x_n \rangle^2_{\tilde{x}_n}}{2 \Delta t}
	- \frac{\kB T}{\gamma} \right] + \kB T ( 1 - \bmxi_n^2 ) + o(\Delta t)
\end{equation}
for $z = \tilde{x}_n$.
As $\bmxi_n^2$ converges to $(dB_t)^2 / dt = 1$ in the limit $\Delta t \to dt$\cite{Sekimoto2010, Gardiner2009},
the dissipative heat at $t$ and position, $\tilde{x} \equiv x(t) + dx(t)/2$, can be formally expressed as
\begin{equation}\label{eq:7}
	d'Q(t; \tilde{x}) = \gamma [ D_\mathrm{inst}(t) - (D_\mathrm{eq} - D_\mathrm{drift}(\tilde{x}))],
\end{equation}
where the three diffusion coefficients are defined as
\begin{equation}\label{eq:8}
	D_\mathrm{inst}(t) \equiv \frac{(dx(t))^2}{2 dt}, \quad
	D_\mathrm{eq} \equiv \frac{\kB T}{\gamma}, \quad
	D_\mathrm{drift}(\tilde{x}) \equiv \frac{\langle dx(t) \rangle^2_{\tilde{x}}}{2 dt}.
\end{equation}
We refer to these three terms as the instantaneous, equilibrium, and drift diffusion coefficients, respectively.
Equation~(\ref{eq:7}) clearly shows that the IDC is intimately related to the dissipative heat, according to the difference, $D_\mathrm{eq} - D_\mathrm{drift}(\tilde{x})$, in NESSs.
Note that Eq.~(\ref{eq:7}) is derived in the limit $\Delta t \to dt$.
Unfortunately, in experiments with a finite time resolution, $\Delta t$, we cannot reach this limit.
Instead, Eq.~(\ref{eq:6}) is practical.
However, the influence of noise with finite order, $\kB T (1 - \bmxi_n^2)$, still remains.

Considering the heat dissipation rate, $J_\mathrm{out} \equiv \frac{1}{N \Delta t} \sum_{n=0}^{N-1} \Delta \widetilde{Q}_n$,
where the measurable dissipative heat is defined as
\begin{equation}
	\Delta \widetilde{Q}_n = \gamma \left[
	\frac{(\Delta x_n)^2}{2 \Delta t} + \frac{\langle \Delta x_n \rangle^2_{\tilde{x}_n}}{2 \Delta t}
	- \frac{\kB T}{\gamma} \right],
\end{equation}
we can remove the finite effect of noise. This is because $\langle \bmxi_n^2 \rangle = \lim_{N \to \infty}\frac{1}{N \Delta t} \sum_{n=0}^{N-1} \bmxi_n^2 = 1$, and $J_\mathrm{out}$ converges to the same limit, $\frac{1}{N \Delta t} \sum_{n=0}^{N-1} \Delta Q_n$.
This means that we can evaluate the heat dissipation rate from single-trajectory data only  without measuring response.
The key to this evaluation is to calculate $\langle \Delta x_n \rangle_{\tilde{x}_n}$ with a high space resolution by using as much data as possible \cite{Comment3}.
Meanwhile, the friction coefficient, $\gamma$, can be easily calculated from trajectories \cite{Toyabe2012}.

\begin{figure}[t]
	\includegraphics[width=0.333\linewidth]{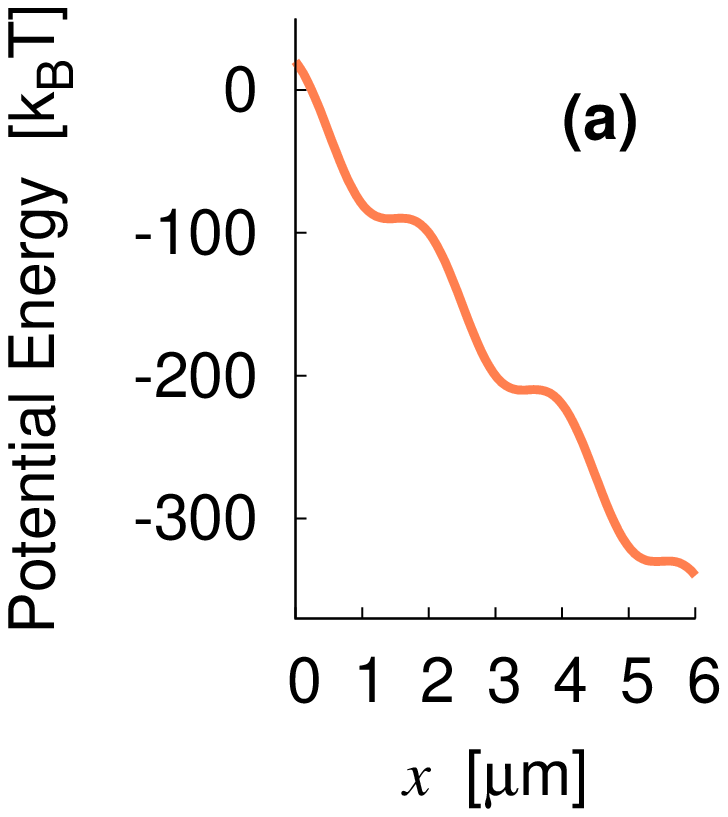}\includegraphics[width=0.333\linewidth]{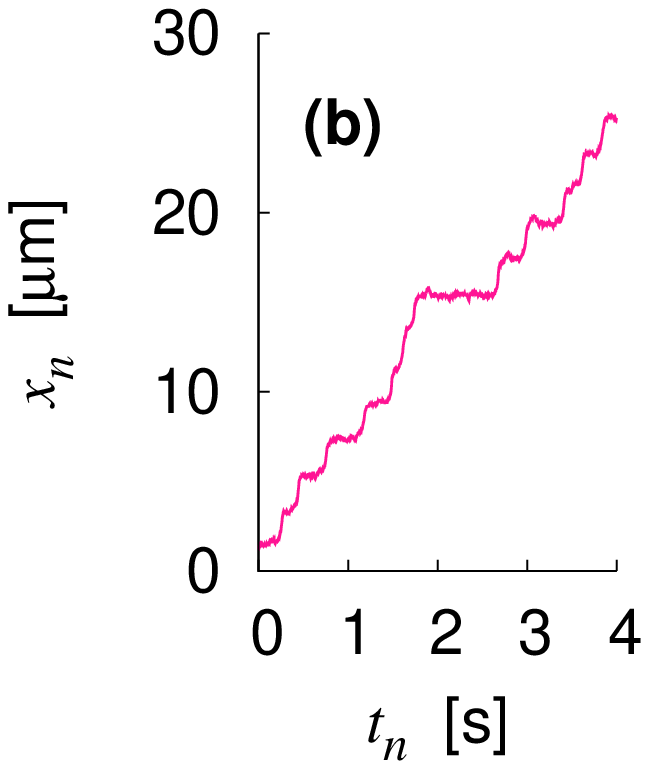}\includegraphics[width=0.333\linewidth]{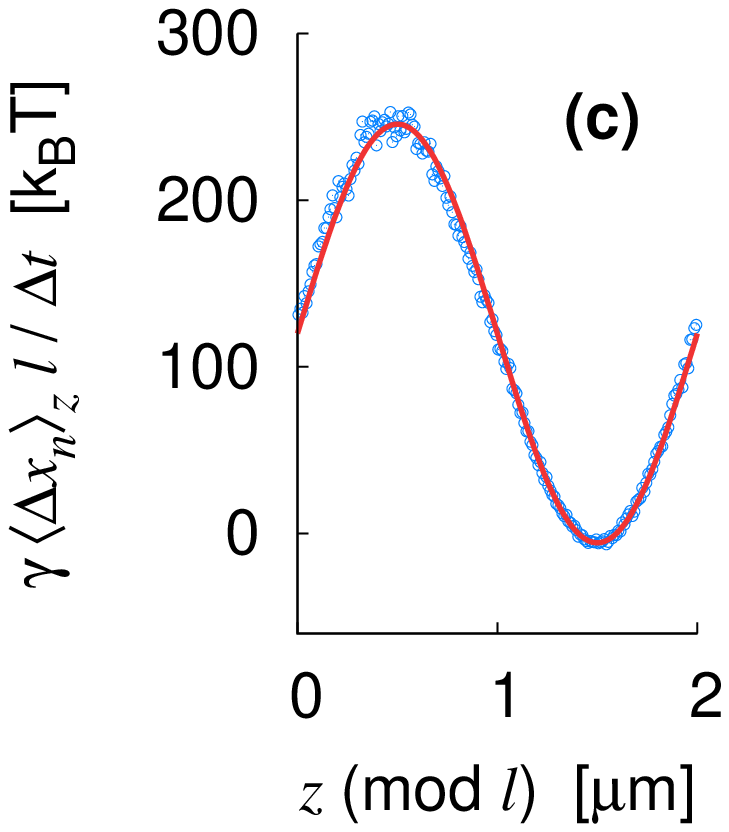}\\
	\includegraphics[width=0.5\linewidth]{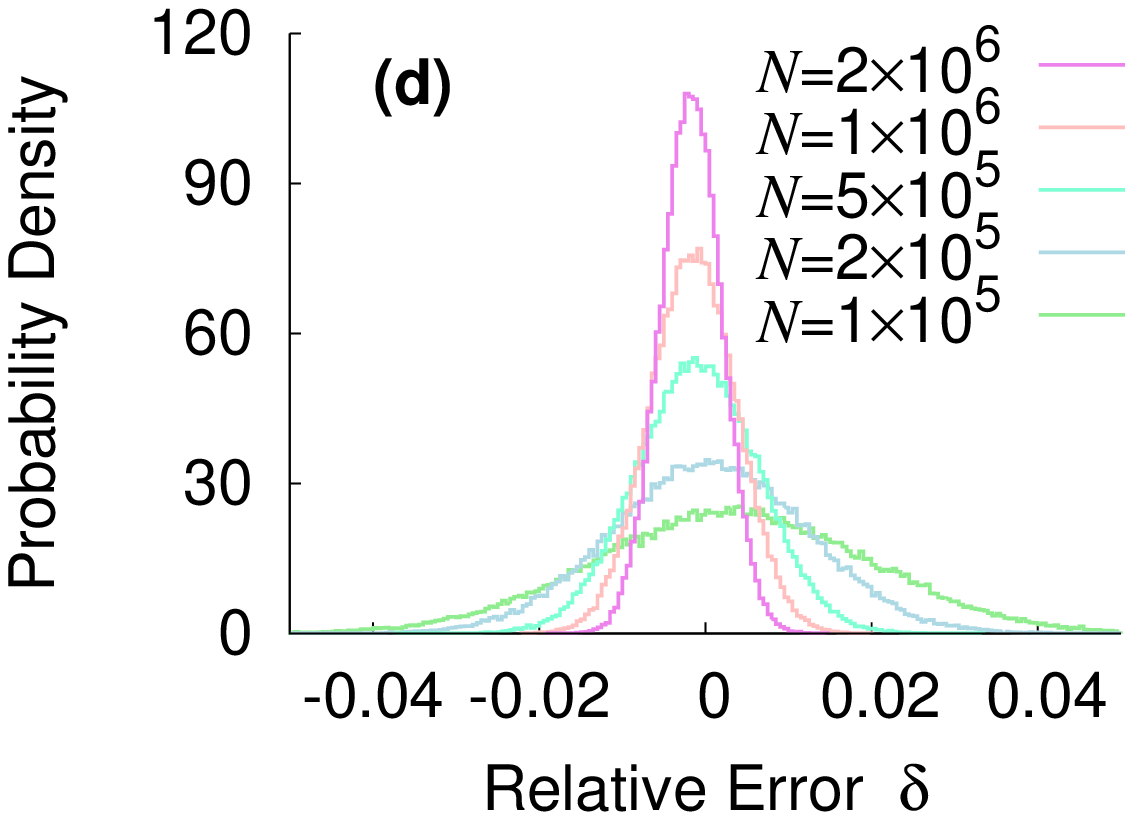}\includegraphics[width=0.5\linewidth]{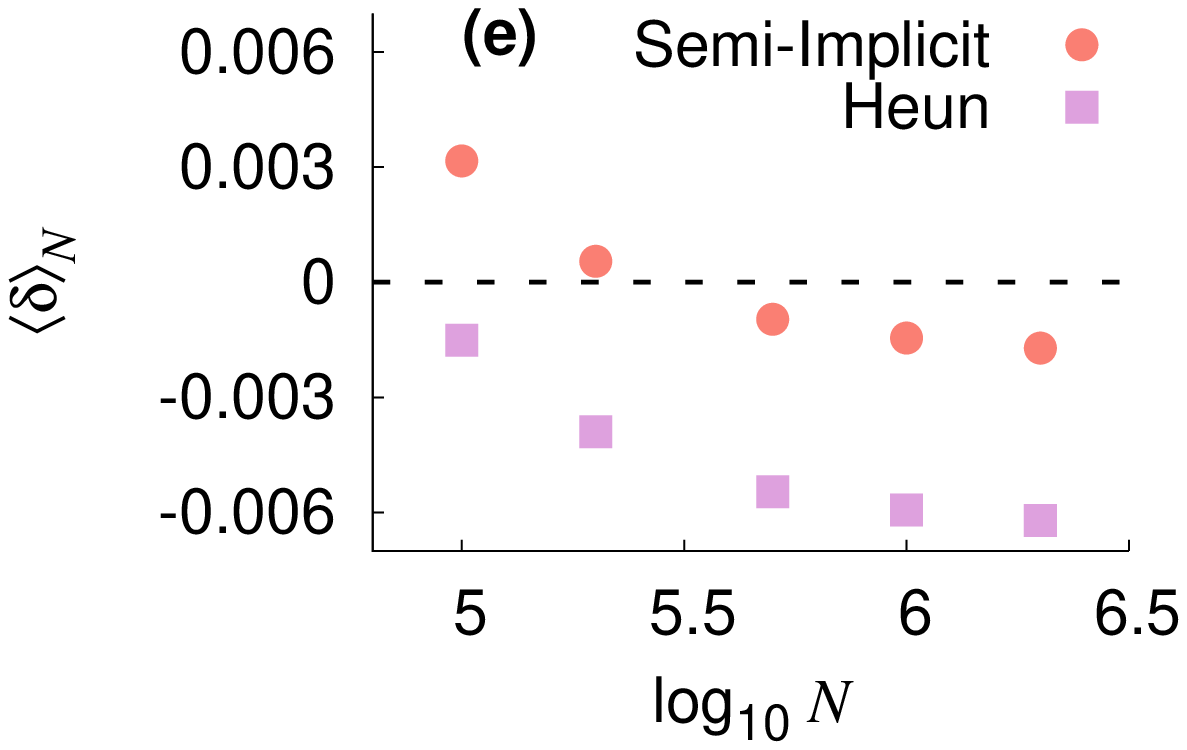}
	\caption{
	(Color online)
	(a) Graph of energy potential, $U(x)  =20 \kB T \cos\left( 2 \pi x / l \right) - F x$, with force, $F = 120 \kB T / l$ N, and periodic length, $l = 2\ \mu$m.
	The unit of energy is $\kB T$.
	(b) Numerical trajectory for parameters: $\gamma = 1.0 \times 10^{-8}$ kg/s, $T = 300$ K, and $\Delta t = 0.001$ s.
	(c) Points indicate $\gamma \langle \Delta x_n \rangle_z l / \Delta t$ for single-trajectory data with $N=10^6$ within each bin, where the space resolution is 10 nm.
	The solid curve represents the theoretical plot of $- U'(z) l / (\kB T) = 40 \pi \sin(2 \pi z / l) + 120$ $\kB T$.
	(d) Probability densities of the relative error, $\delta \equiv (J_\mathrm{out} - J_\mathrm{in}) / J_\mathrm{in}$, for $N=1 \times 10^5, \, 2 \times 10^5, \, 5 \times 10^5, \, 1 \times 10^6, \, 2 \times 10^6$.
	Each ensemble consists of $10^5$ trajectories.
	(e) Average relative error of $N$, $\langle \delta \rangle_N$.
	Filled circles and squares correspond to the semi-implicit and the Heun algorithm, respectively.
	}
	\label{fig:1}
\end{figure}

Finally, we give a numerical demonstration to evaluate the heat dissipation rate.
As an example, let us consider a tilted periodic potential, $U(x) = 20 k_\mathrm{B} T \cos(2 \pi x /l) - F x / l$, with force, $F = 120 \kB T / l$ N, and periodic length, $l = 2\ \mu$m.
Figure~\ref{fig:1}(a) shows the graph of this potential, where the unit of energy is $\kB T$.
Unless the numerical calculation of the integral of the Langevin equation~(Eq. (\ref{eq:1})) is in the form of a Stratonovich-type SDE, excess energy is generated in contradiction to the law of energy conservation \cite{Sekimoto2010}.
Here, we computed the integral using the following semi-implicit algorithm \cite{Gardiner2009}:
\begin{equation}
	\tilde{x}_n = x_n - \frac{1}{2} \frac{U'(\tilde{x}_n)}{\gamma} \Delta t
	+ \frac{1}{2} \sqrt{\frac{2 k_\mathrm{B} T \Delta t}{\gamma}} \bm{\xi}_n, \quad
	x_{n+1} = 2 \tilde{x}_n - x_n.
\end{equation}
A numerical trajectory for the parameters, $\gamma = 1.0 \times 10^{-8}$ kg/s, $T = 300$ K, and $\Delta t = 0.001$ s, is shown in Fig.~\ref{fig:1}(b).
Here, $\langle \Delta x_n \rangle_ z$ was calculated within each bin, where the space resolution was 10 nm, for a single trajectory with $N=10^6$.
Figure~\ref{fig:1}(c) indicates that the calculated $\langle \Delta x_n \rangle_z$ (points) show good agreement with the theoretical curve (solid curve).
In order to determine the accuracy of our method of evaluating the heat dissipation rate, we considered the energy rate balance between the output, $J_\mathrm{out}$, and the input, which is defined as $J_\mathrm{in} \equiv F \frac{1}{N \Delta t}\sum_{n=0}^{N-1}\Delta x_n$.
For each trajectory, we calculated the relative error, $\delta \equiv (J_\mathrm{out} - J_\mathrm{in}) / J_\mathrm{in}$.
The probability densities of $\delta$ for $N=1 \times 10^5, \, 2 \times 10^5, \, 5 \times 10^5, \, 1 \times 10^6, \, 2 \times 10^6$ are shown in Fig.~\ref{fig:1}(d).
Each ensemble consists of $10^5$ trajectories, and the ensemble average of $N$, $\langle \delta  \rangle_N$, is shown in Fig.~\ref{fig:1}(e).
In our simulations, the average didn't converge to just zero for not only the semi-implicit algorithm but also the Heun algorithm\cite{Sekimoto2010}.
The accurate convergence might strongly depend on the algorithm for solving the SDE.

The development of stochastic energetics by Sekimoto has changed researchers' perspective on the fluctuating world.
For example, diffusion phenomena in molecular motors or living cells can be analyzed in terms of the stochastic energetics based on single-trajectory data only \cite{Toyabe2012, ShinkaiTogashi2014}, and the dissipative heat can be related to some form of nonequilibrium biological activity\cite{Mizuno2007}.
The method presented in this note has potential application in the analysis of single-molecule trajectories in living cells, which will facilitate the discovery of the relationship between molecular function and cell environment.

\begin{acknowledgment}
This work was inspired from communications with Shoichi Toyabe, who sent comments following the publication of our previous paper \cite{ShinkaiTogashi2014}, for which I am sincerely grateful. I also gratefully acknowledge Akinori Awazu and Yuichi Togashi for helpful discussions.
This work was supported by MEXT, Japan (Platform for Dynamic Approaches to Living System; KAKENHI 23115007).
\end{acknowledgment}

\end{document}